\documentclass[aps,twocolumn,showpacs]{revtex4}
\usepackage{color} 
\usepackage[english]{babel}
\usepackage{graphicx}
\usepackage{dcolumn}

\def\beq#1{\begin{equation}\label{#1}}
\def\eeq{\end{equation}}

\begin{document}
\title{Attosecond pulse production using resonantly-enhanced high-order harmonics}
\author{ V.V. Strelkov\footnote {strelkov.v@gmail.com}
 }
\affiliation{A. M. Prokhorov General Physics Institute of the RAS, Moscow 119991,  Russia \\
Moscow Institute of Physics and Technology (State University), 141700 Dolgoprudny, Moscow Region, Russia }

\begin{abstract}
\noindent 
We study theoretically the effect of the giant resonance in Xe on the phase difference between the consecutive high order resonantly-enhanced harmonics and calculate the duration of the attosecond pulses produced by these harmonics.  For certain conditions resonantly-induced dephasing compensates the phase difference which is intrinsic for the off-resonance harmonics. We find these conditions analytically and compare them with the numerical results. This harmonic synchronization allows attosecond pulse shortening in conjunction with the resonance-induced intensity increase by more than an order of magnitude; the latter enhancement relaxes the requirements for the UV filtering needed for the attosecond pulse production.   Using a two-color driving field allows further increase of the intensity. In particular, a caustic-like feature in the harmonic spectrum leads to the generation efficiency growth up to two orders of magnitude, however accompanied by an elongation of the XUV pulse.
\end{abstract}
\pacs{
  42.65.Ky            
  32.80.Rm            
}
\maketitle
\noindent

{ Attosecond pulse production using high order harmonics generated by intense laser field~\cite{Paul, Tzallas} is essentially based on the phase-locking of the harmonics. This phase-locking is well understood~\cite{Antoine,Salieres2001} for the case when there are no resonances affecting the process. However, recently much attention has been paid to the role of resonances in high harmonic generation (HHG) in gases~\cite{Gilbertson,Xe-Kr, caustic2, Rothhardt} and plasma plumes~\cite{Ganeev0,Haessler-atto,Rosenthal} (for a review of earlier studies see also~\cite{Ganeev1,Ganeev2}). It was shown that when the high harmonic frequency is close to the transition to an excited quasi-stable state of the generating particle the harmonic can be much more intense than the off-resonant ones. For the HHG in plasma plumes such enhancement can be as high as an order of magnitude of even more.} The XUV generation efficiency enhancement due to the giant resonance in Xe was predicted in~\cite{Frolov} and observed in~\cite{Xe-Kr,caustic2}. Namely, the XUV near 100 eV in the spectral region of about 20 eV is more intense than the lower-frequency XUV, and the enhancement near the center of the resonance is approximately an order of magnitude.  

The broadband resonant enhancement potentially allows generating attosecond pulses using resonant harmonics. This approach is interesting not only because of the higher generation efficiency of the resonant HH, but also because it essentially reduces the requirements for harmonic filtering (the resonant region is naturally standing out). However, the phase-locking of resonant HH differs from the one of the non-resonant HH~\cite{Haessler-atto}, so the attosecond pulse production in the former case {is not straightforward}. In this Rapid Communication we investigate this aspect of resonant HHG both numerically and analytically. We study the effect of the resonance on the phase difference between the neighbor harmonics and calculate the duration of the attosecond pulses produced by resonant harmonics. 

The time-dependent three-dimensional Schr\"odinger equation (TDSE) is solved numerically for a single-active electron atom in an external laser field. The method of the numerical TDSE solution is described in~\cite{JPB_num}.The model atomic potential is (atomic units are used throughout):
\beq{model_potential} 
V(r)= - a_0 \exp \left( -\frac{r^2}{b_0^2}\right)+a_1 \exp \left( -\frac{(r-r_0)^2}{b_1^2}\right)
\eeq 

The first term is the binding potential of the atomic core, and the second one is the barrier providing a bound quasi-static state with positive energy. The potential is similar to the one used earlier in the resonant HHG calculations~\cite{Strelkov_4step, Tudorovskaya, HHG_Fano}. Moreover, a double-barrier effective potential was found in~\cite{Kapoor}  describing autoionization in time-dependent density-functional theory. 

In the potentials that we used in~\cite{Strelkov_4step, HHG_Fano} the first term is a soft-Coulomb potential, whereas in the potential~(\ref{model_potential}) it is Gaussian. Such potential does not have the Coulomb 'tails' and thus does not support Rydberg states; however, it provides more freedom to simulate the properties of the desired atom. Choosing the parameters $a_0$, $a_1$, $b_0$, $b_1$, and $r_0$  of the model potential we reproduce the ionization energy of Xe atom,  {frequency and width of the giant resonance  so that the frequency and width of the resonantly enhanced region in the calculated HH spectrum is close to those observed } in Ref.~\cite{Xe-Kr}. The parameters of the potential used in our calculations are $a_0=7.1$, $b_0=1.0$,  $a_1=4.5$, $b_1=0.5$ and $r_0=1.23$ a.u..  Throughout this paper we simulate only the shortest electronic trajectory {(if not specially stated otherwise)}, suppressing the others with properly defined absorbing region in the numerical box, as it was done in~\cite{harmonic_ellipticity_PRA}.
\begin{figure}  
\centering
\includegraphics [width=0.8\columnwidth] {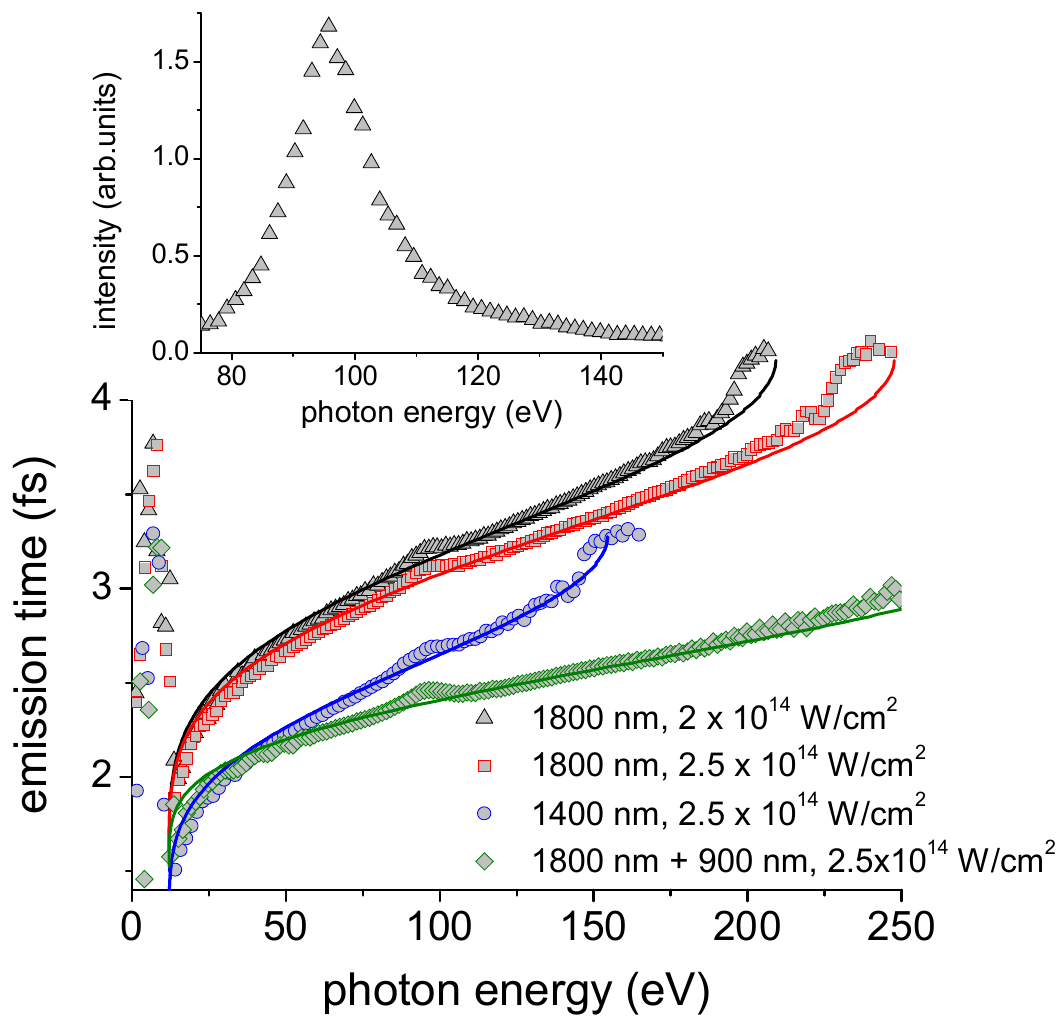}
\caption{(color online) Emission time calculated via numerical TDSE solution (symbols) and the estimates of this time as a classical return time of the electron in the simple-man model (solid lines). The driving laser intensities and wavelengths are  shown in the graph; the two-color field is given by eq.~(\ref{two-color}) with $\alpha=1$ and $\phi=\pi/2$. The inset shows harmonic intensities near the resonance.}
\label{fig1}
\end{figure}

The laser field intensity is switched on smoothly during 4 optical cycles, then it is constant during 4 cycles, and then decreases during 4 cycles; the shape of the laser pulse is described in~\cite{Strelkov_2006}.
 {We are using either a single-color driving field or a two-color one. The two-color field is given by  }
\beq{two-color}
E(t)=E_0 f(t) [\exp(-i \omega_l t)+\sqrt{\alpha} \exp(-i 2 \omega_l t +i \phi)]+ c.c.
\eeq
 {where $E_0$ is the amplitude of the main component of the laser field, $\omega_l$ is its frequency,  $f(t)$ is the slowly-varying envelope, $\alpha$ is the ratio of the intensities of the driving field components, $\phi$ is the relative phase.}

 We calculate the harmonic spectrum and find the spectral phase differences between the consecutive harmonics. Then we calculate the emission times $t_e (q \omega_l) =(\varphi_{q+1}- \varphi_{q-1})/(2 \omega_{l})$, where $q$ is an even number, $\varphi_{q \pm 1}$ are the harmonic spectral phases; $t_e$ characterizes the time instant when an attosecond pulse formed by a group of harmonics with the central frequency $q \omega_l$ is emitted~\cite{Mairesse_2003}. The results are shown in Fig~\ref{fig1} for different laser intensities and wavelengths together with the classical electronic return times $t_r (q \omega_l)$ calculated within the simple-man approach~\cite{simple-man1,simple-man2}. Times  $t_e$ and $t_r$ in general are close to each other~\cite{Antoine}; the agreement is very good under the conditions of our calculations which are well within the tunneling regime of ionization. However, the pronounced deviation from the classical prediction can be seen in the cut-off region and in the resonant region. The phase-locking of the harmonics near the cut-off was studied recently in details in~\cite{Khokhlova_2016}; in this paper we will study the harmonic phase-locking in the resonant region. 
\begin{figure}  
\centering
\includegraphics [width=0.7\columnwidth] {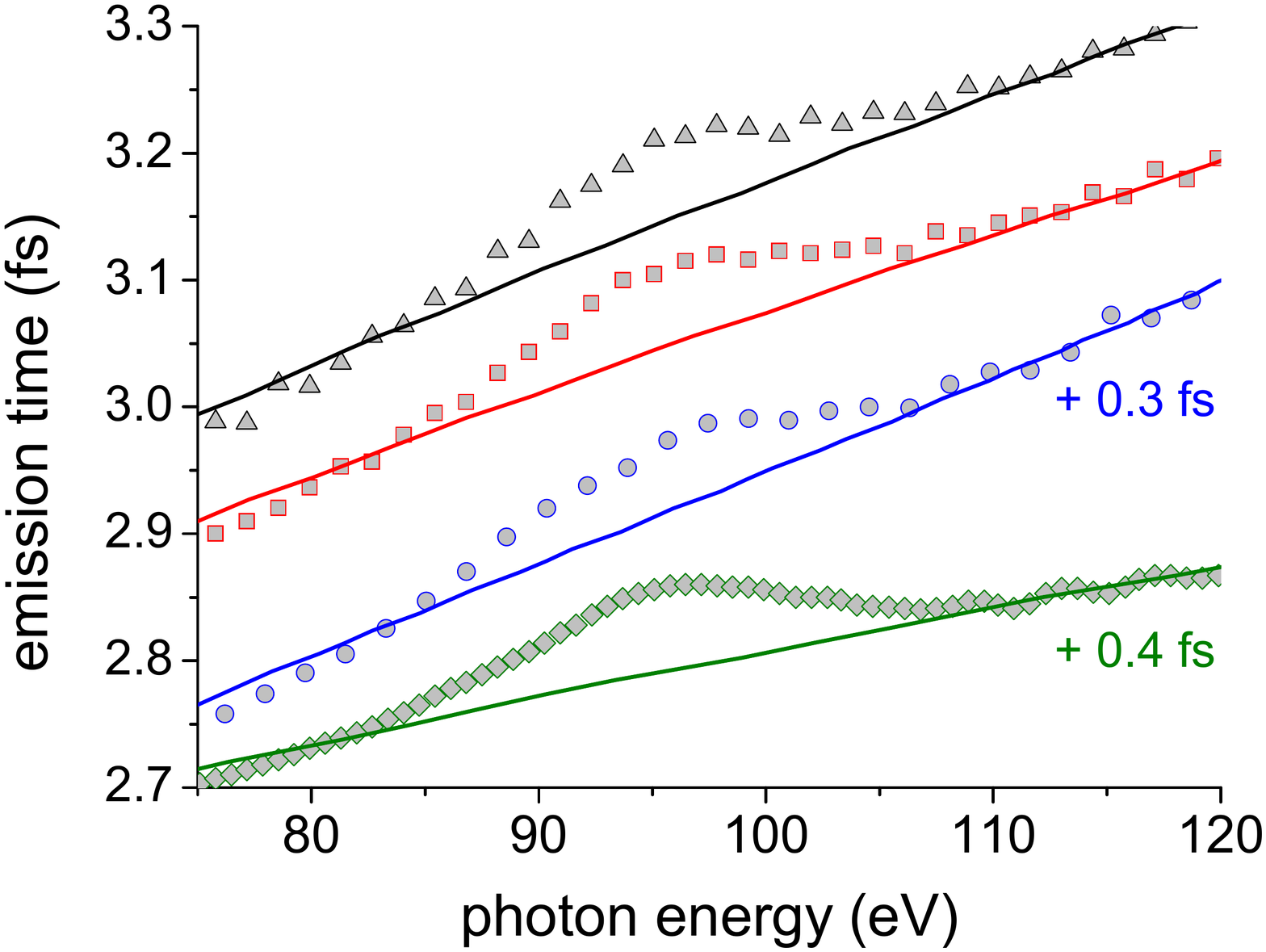}
\caption{(color online) Close-up of the harmonic emission times near the resonance. The fields' parameters are the same as in Fig.~\ref{fig1}.}
\label{fig2}
\end{figure}

The inset in Fig.~\ref{fig1} shows the harmonic spectrum near the resonance. The group of harmonics around approx. 96~eV are enhanced, and the harmonic intensity in the center of the resonance is about an order of magnitude higher than far from it. The width of the resonantly-enhanced harmonic group is 17~eV (FWHM).

Fig.~\ref{fig2} shows the emission times for these harmonics. To calculate these emission times we used harmonic phases averaged over 10 laser intensities in the vicinity (within $\pm 2 \%$) of the intensity presented in the figure; this is done to reduce the numerical noise. We can see that the emission time is affected by the resonance: the resonant harmonics are emitted {\it later} than they would be emitted in the absence of the resonance. This result is in agreement with the published experimental results for HHG in Sn$^+$~\cite{Haessler-atto}, as well as analytical and numerical studies~\cite{Tudorovskaya,HHG_Fano}.  The found delay time for the harmonics near the center of the resonant line (68 as) is close to the lifetime of the quasi-stable state (77 as). This result can be well understood within the four-step model of the resonant HHG~\cite{Strelkov_4step}: the resonant XUV emission is delayed with respect to the non-resonant one, and the delay time is the time which the system stays in the quasi-stable state after rescattering.

The resonant-induced delay of the XUV emission smoothly decreases with the increase of the detuning from the resonance. So, in the spectral region above the resonance this 'resonant attochirp' can compensate the usual 'free-motion attochirp' (the one caused by the free electronic motion before rescattering).  Let us estimate both these chirps and their influence on the attosecond pulse duration.

To do this we consider the chirped Gaussian attopulse:
\beq{pulse} 
\begin{array}{l}
F(t)=\exp(-i \Omega t) \int_{-\infty}^{+\infty} \exp\left( -2 \ln(2) \left( \frac{\omega'}{\Delta \omega}\right)^2 \right) \\
\exp \left( i \frac{K}{2} \omega'^2 \right) \exp\{-i \omega' (t-t_r(\Omega))\} d \omega'
\end{array}
\eeq
where $\Omega$ is the central frequency of the pulse and $t_r(\Omega)$ is the emission time of the pulse. Let us assume that the chirp of the pulse is only due to the variation of the emission frequency described by the simple-man model (below we denote this chirp as the 'free-motion-induced attochirp'). Thus the derivative of the spectral phase ($\varphi \equiv \frac{K}{2} \omega'^2+t_r(\Omega) \omega' $) over the frequency $\omega' $ is the classical electronic return time $t_r(\omega')$. So $K=\partial t_r / \partial \omega'$. From Fig.~\ref{fig1} we can see that $t_r$ is an almost linear function of $\omega'$ except for the lowest and the highest part of the plateau. This linear approximation can be found from the solution of the Newton's equation for the electron in the simple-man model. We find that approximately 
\beq{K}
K=1/(2 \omega_l U_p) 
\eeq 
where $U_p$ is the ponderomotive energy. The duration of the chirped pulse~(\ref{pulse}) depends on its spectral width $ \Delta \omega$. The shortest duration is achieved when $
\Delta \omega = \sqrt{4 \ln(2)/K}
$.
Substituting~(\ref{K}) in the latter equation we find that 
\beq{Dw1}
\Delta \omega = 2 \sqrt{2 \ln (2) U_p \omega_l}
\eeq
The duration (FWHM of intensity) of this pulse is
\beq{duration}
\tau = 2 \sqrt{ \ln (2)/(U_p \omega_l)}
\eeq
Estimates~(\ref{Dw1}) and~(\ref{duration}) agree very well with the numerical TDSE calculations for the off-resonant harmonics. Similar estimate of the shortest attosecond pulse duration was found in~\cite{Platonenko_1997}. Note that equation~(\ref{duration}) shows that  $\tau \propto \sqrt{\omega_l /I}$ where $I$ is the laser intensity.  { Since the maximum laser intensity is practically limited by the target ionization, this equation shows that the minimum attopulse duration decreases with the laser frequency decrease.} 

The 'resonantly-induced attochirp' can be estimated taking into account the delay in the resonant XUV emission. As we discussed above, this delay is the lifetime of the resonance (denoted below as $\Delta t $) for the XUV in the center of the resonance, and it vanishes within the width of the resonance $\Gamma$. So the resonantly-induced attochirp is $K_{res}= - \Delta t / \Gamma$. Having in mind that $\Delta t=1/\Gamma$ we find that 
$
K_{res}=-\Gamma^{-2}
$.
From this equation and equation~(\ref{K}) we find that $K=-K_{res}$ for
\beq{result}
\Gamma=\sqrt{2 U_p \omega_l}
\eeq
\begin{figure}  
\centering
\includegraphics [width=0.8\columnwidth] {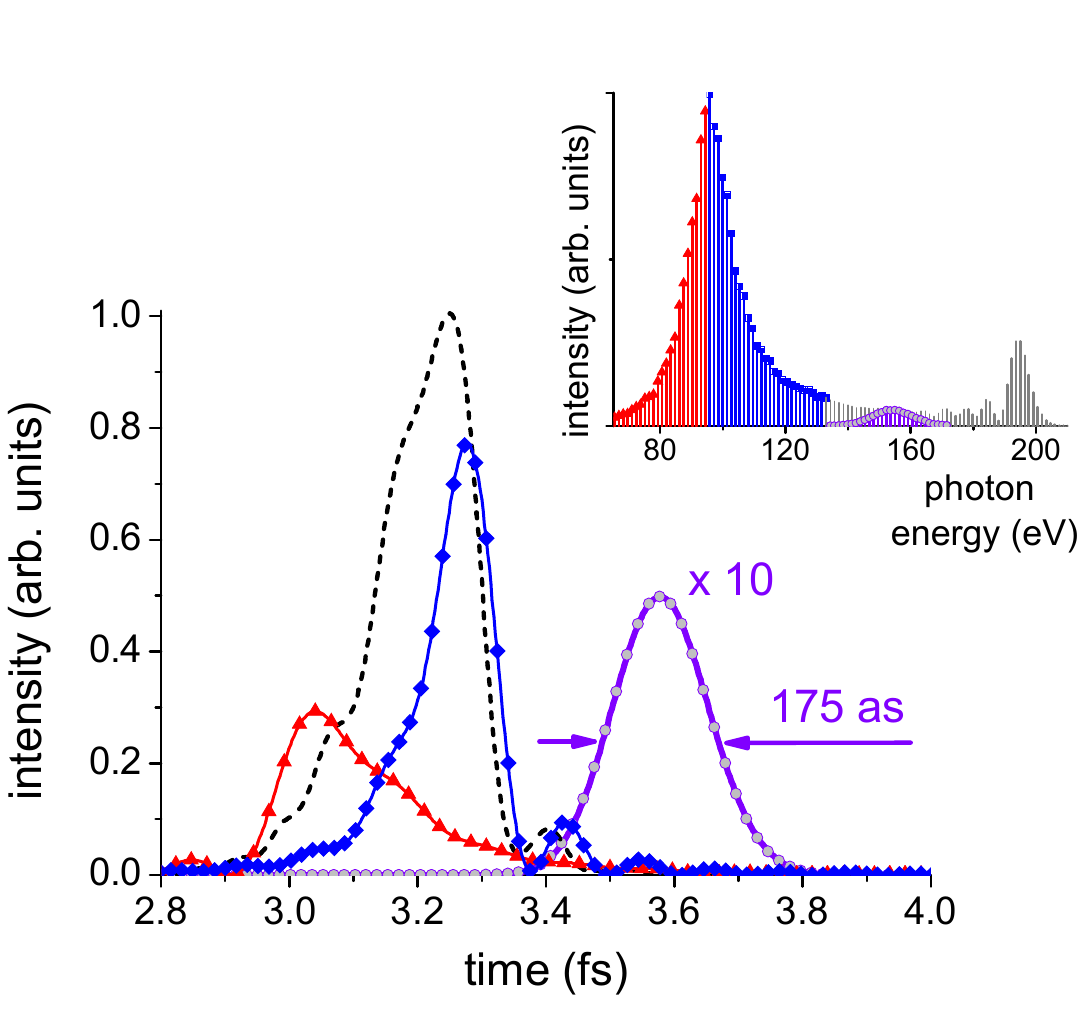}
\caption{(color online). The attosecond pulses calculated via equation~(\ref{attosecond_pulses}) using resonantly-enhanced harmonics below the center of the resonance (red lines with triangles), above the center of the resonance  (blue lines with diamonds), all resonantly-enhanced harmonics (dashed black line) and the group of off-resonant harmonics chosen to minimize the attopulse duration (violet line with circles), the latter attopulse is multiplied by 10; see text for more details. The inset shows the harmonic spectrum; the spectral regions used to calculate the attosecond pulses are highlighted with the corresponding colors. The laser wavelength is 1800 nm and the intensity is $2 \times 10^{14}$ W/cm$^2$.}
\label{fig3}
\end{figure}
\begin{figure}  
\centering
\includegraphics [width=0.7\columnwidth] {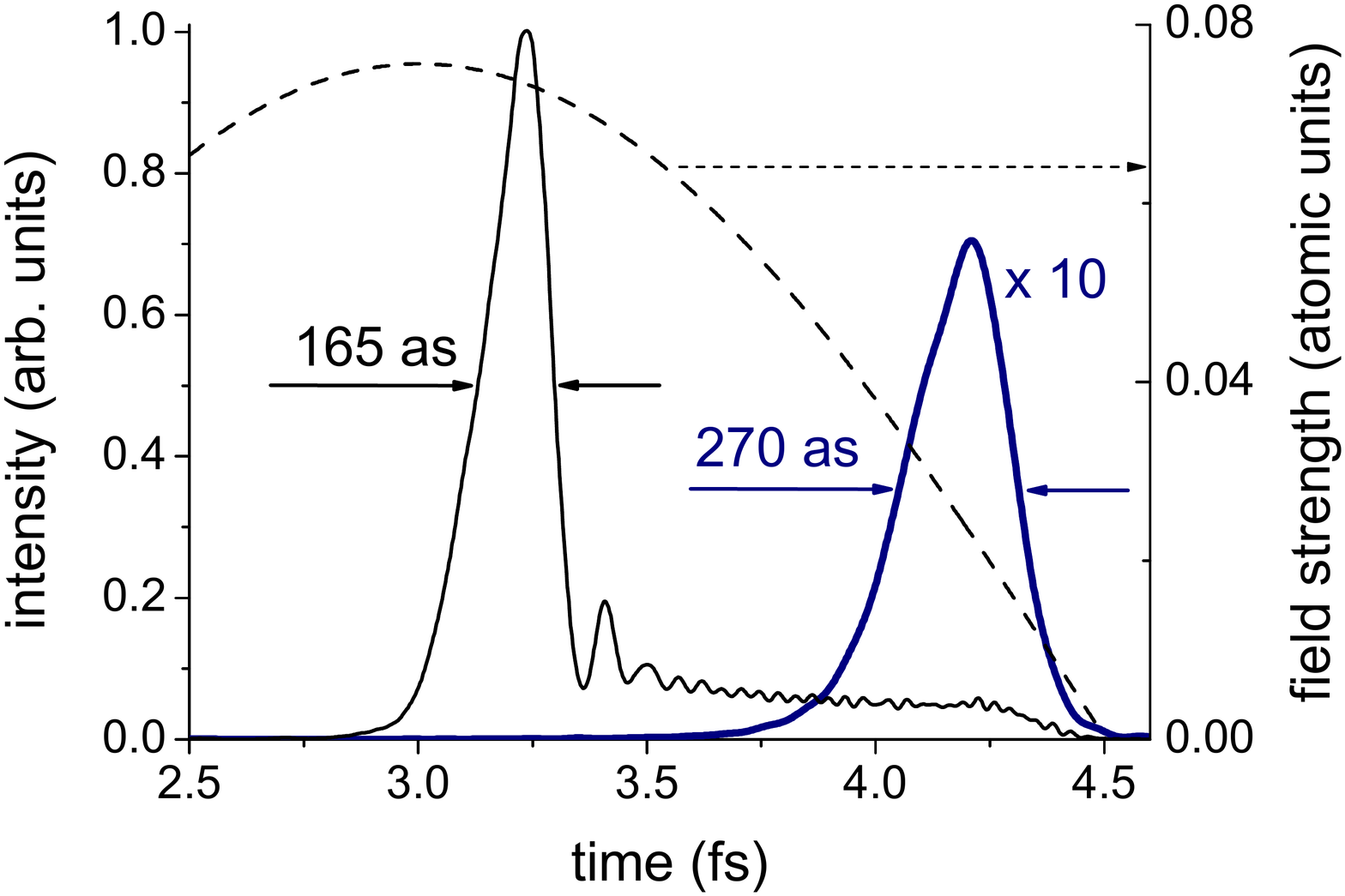}
\caption{(color online). The attosecond pulses formed by all harmonics higher than 60 eV (thin black line) and higher than 190 eV (thick navy line); the latter is multiplied by 10. The instantaneous strength of the laser field is shown with dashed line. The laser parameters are the same as in Fig.~\ref{fig3}}
\label{fig4}
\end{figure}

The conditions of our calculations {for the single-color driving field} were chosen so that the latter equation is approximately satisfied: we can see that in Fig.~\ref{fig2} the free-motion-induced attochirp is compensated with the resonantly-induced one, so the group of harmonics above the central frequency of the resonance have approximately the same phases. {In contrast to this, the parameters of the two-color field used in our calculations lead to smaller free-motion induced attochirp, so the resonantly-induced one dominates.}

 In Figs.~\ref{fig3} and~\ref{fig4} we show the attosecond pulses calculated using XUV from different spectral regions. Namely, using the complex amplitudes of the microscopic response $d(\omega)$ calculated via numerical TDSE solution we find the XUV intensity:
\beq{attosecond_pulses} 
I(t)= \left | \int_{\omega=-\infty}^{\infty} M(\omega) d(\omega) \exp(-i \omega t) d \omega \right | ^2   
\eeq 
where the used spectrum mask $M(\omega)$ is either a Gaussian $M_{G}(\omega)=\exp\{-2 \ln(2)((\omega-\Omega)/\Delta \omega)^2\}$ or a step-like function: $M_{step}(\omega)=\theta(\omega-\omega_{low})\theta(\omega_{high}-\omega)$.
In Fig.~\ref{fig3} we present the attopulses formed by resonant harmonics below the resonance  (calculated using $M_{step}$ with $\omega_{low}=60$eV and $\omega_{high}=96$eV), above it ($\omega_{low}=96$eV and $\omega_{high}=130$eV), and all the resonant harmonics ($\omega_{low}=60$eV and $\omega_{high}=130$eV). We can see that the attosecond pulse formed by the harmonics above the resonance is much shorter than the one formed by those below the resonance. This is because the above-resonant harmonics are in phase, whereas those below the resonance have significant phase differences (see Fig.~\ref{fig2}), as it was discussed above. In the same figure we show the attopulse formed by the off-resonance harmonics calculated using $M_{G}$ with the central frequency $\Omega=155$ eV and the width $\Delta \omega$=15.2 eV. The latter is found numerically to minimize the pulse duration; the found width and duration are very  close to the predictions of eq.~(\ref{Dw1}) and~(\ref{duration}), respectively. We can see that this pulse is slightly longer and much weaker than the one formed by the above-resonance harmonics. Moreover, if all the resonant harmonics are used to produce the attopulse, its duration does not increase dramatically, see dashed line in the Fig.~\ref{fig3}.       

{Metal foils or multilayer mirrors are usually applied as spectral filters~\cite{Paul,Tzallas,filter,Goulielmakis} to obtain attosecond pulses; such filter, in particular, can transmit well all UV higher than certain frequency.  To simulate the attosecond pulses obtained with such filter we use in equation~(\ref{attosecond_pulses}) the step-like mask with $\omega_{high}$ which is much higher than the cut-off frequency. The results obtained using $\omega_{low}$ well-below and well-above the resonance are shown in  Fig.~\ref{fig4}.} Again, in the latter case this number is chosen to minimize the duration of the attosecond pulse. In spite of this optimization, we can see that this attopulse using off-resonant harmonics is longer and more than an order of magnitude weaker than the resonant one.  Note that the parameters of the attosecond pulse formed by the resonant harmonics are not very sensitive to $\omega_{low}$ as long as it is well-below resonance; this is natural because the off-resonant harmonics are much weaker than the resonant ones. This means that practically there is much freedom  in choosing such filter as long as it transmits the resonant harmonics. Moreover, if the absorption edge  of the filter is far from the resonance, the filter dispersion (which is usually pronounced only in the vicinity of the absorption edge) would not affect the attosecond pulse duration. So the duration of 165~as found in our calculations is close to the one which can be experimentally obtained using harmonics enhanced by the giant resonance in Xe.
\begin{figure}  
\centering
\includegraphics [width=0.8\columnwidth] {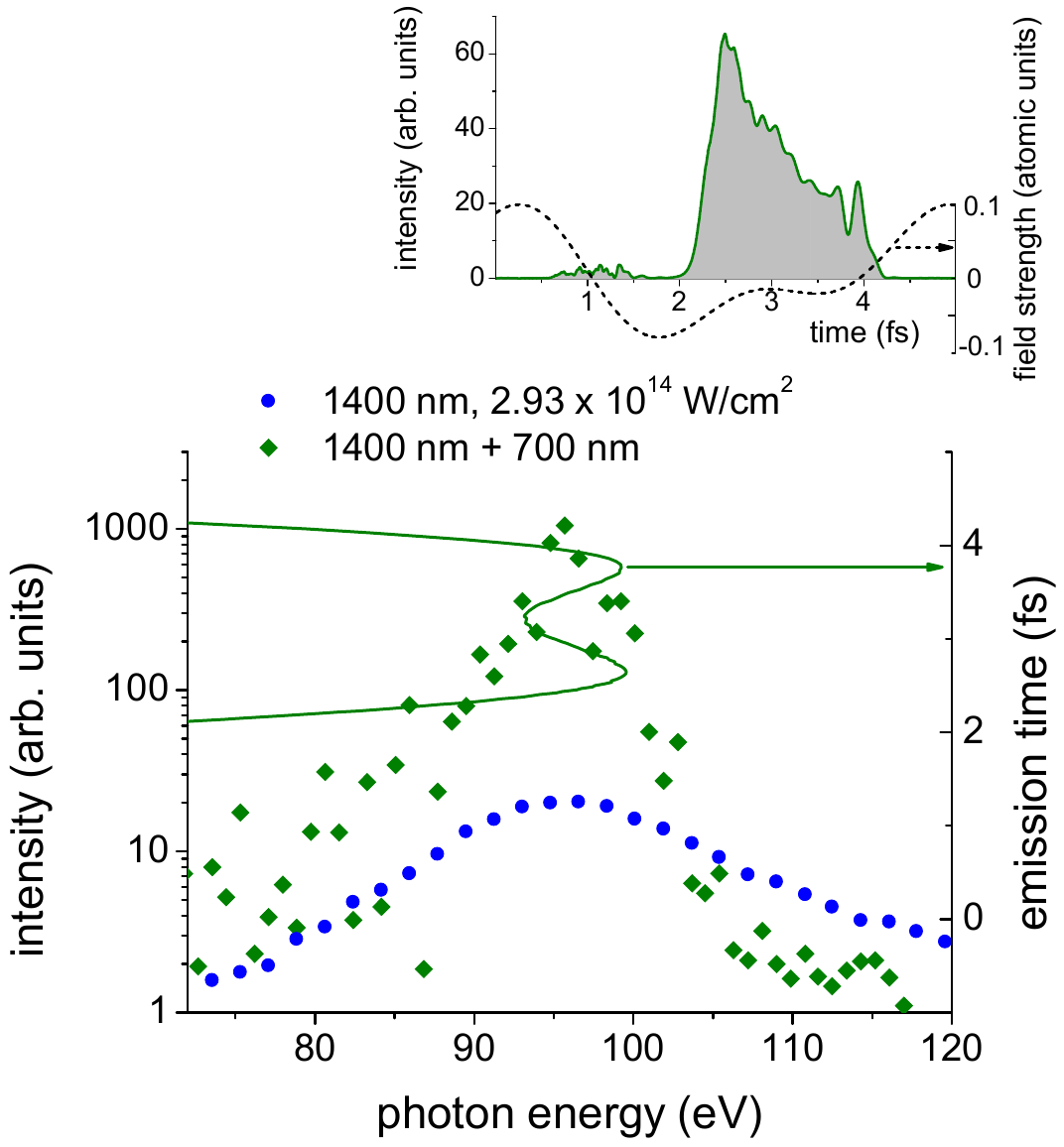}
\caption{(color online). Harmonic intensities generated in single-color (blue circles) and two-color (green diamonds) fields, see text for more details, and the classical return time for the two-color case. The inset shows the driving field strength (dotted line) and the resonant XUV pulse intensity (solid line) as functions of time.}
\label{fig5}
\end{figure}

{Making similar calculations for the HHG by the two-color field with the parameters $\alpha$ and $\phi$ considered above we find that the attopulse formed by above-resonant XUV is longer, and the one formed by below-resonant XUV is shorter than in the single-color field. This is the result of the attochirp behavior shown in Fig.~\ref{fig2}: the absolute value of the total attochirp above the resonance is higher in the two-color case, below the resonance the relation is reversed. Note that this leads to smaller dephasing between the harmonics near the very center of the resonance. Since these harmonics are the most intense ones, this results in even shorter attosecond pulse than in the single-color field. Namely, the attosecond pulses formed by all XUV with frequency higher than $\omega_{low}=60$eV can be as short as 105 fs in the two-color case.}

{As it was shown both theoretically~\cite{JPB_num} and experimentally~\cite{Colosimo, Shiner} the harmonic yield rapidly decreases with the decrease of the driving wavelength. So the perspective of the generation efficiency increase using two-color field~\cite{Eichmann,Kim, Emelina} is especially important for the middle-infrared drivers considered here. To achieve maximum resonant harmonic intensity in the two-color field we chose the parameters $\alpha$ and $\phi$ of the latter so that  the resonant frequency coincides with the caustic-like feature~\cite{caustic1,caustic2} in the dependence of the returning electron energy on the return time. Due to this feature almost all detached electrons return back with the energy close to the one of the quasi-stable state. This leads to a further increase of the resonant harmonic generation efficiency. Fig.~\ref{fig5} shows the results calculated for the fundamental intensity  $2.2 \times 10^{14}$ W/cm$^2$, $\alpha=1/3$, $\phi=1.0$ (here we take into account all the electronic trajectories in the TDSE). We compare the HHG efficiency in the two-color field with that in the single-color field having the intensity equal to the sum of the intensities of the fundamental and the second harmonic in the two-color case. Fig.~\ref{fig5} shows that the gain from using the two-color field with the proper parameters can be about two orders of magnitude. Together with the resonance-induced enhancement this provides the level of conversion which can be interesting for using such harmonics as an efficient source of coherent XUV in the range of 100 eV. However, the inset in Fig.~\ref{fig5} shows that the generation efficiency increase using such caustic-like feature leads to a loss of the attosecond nature  of the emitted XUV: the calculated XUV pulse is of approximately 2 fs duration. This value is close to the time interval when the classical electrons return with energy close to the resonant one.}

{Thus in this paper we find conditions for which the free-motion-induced attochirp can be compensated by the resonantly-induced attochirp, leading to phase synchronization of a group of resonant harmonics. It is shown that attopulses with duration of 165 as can be obtained using resonantly-enhanced harmonics generated in Xe. This duration is smaller than the minimal duration of the attosecond pulse formed by the off-resonant harmonics; it can be further reduced down to almost hundred attoseconds using the two-color driver. Resonant HHG enhancement leads to an increase of the attopulse intensity by more than an order of magnitude and relaxes the requirements for the XUV filtering: only harmonics much lower than the resonance should be suppressed by the filter. Using two-color field with specific parameters providing 'caustic-induced' enhancement of the resonant harmonics provides further (almost two-orders of magnitude) increase of the XUV intensity at the cost of increasing the pulse duration to above 1 fs. }

{Note that the giant resonances are observed also in other atoms, ions and molecules. The detailed investigation of their applicability for attosecond pulse production can be a natural development of the present study. However, our findings, in particular, the estimate of the laser field parameters required for the phase synchronization of the resonant harmonics given by eq.~(\ref{result}) should be applicable for other resonances as well.}

This study was funded by RSF (grant N 16-12-10279).

\end{document}